\documentclass[12pt,a4paper]{article}

\usepackage{graphicx}
\usepackage{dcolumn}
\usepackage{bm}
\usepackage{amssymb}
\usepackage{amscd}
\usepackage{amsmath}
\usepackage{afterpage}
\usepackage{float,times}
\usepackage{subfigure}
\usepackage{rotating}
\usepackage{multirow}
\usepackage{fancyheadings}
\usepackage{epsfig}
\usepackage{theorem}
\usepackage{moreverb}
\usepackage{euscript}
\usepackage{psfrag}
 
\begin{document}

\begin{titlepage}

\noindent
\begin{center}
\vspace*{1cm}

{\large\bf COSMOLOGICAL ZITTERBEWEGUNG} 

\vskip 1cm

{\bf Parth Girdhar}$^{\rm b,}$\footnote{\tt pgir1104@uni.sydney.edu.au} {\bf and
  Archil Kobakhidze}$^{\rm a,b,}$\footnote{\tt archilk@physics.usyd.edu.au}   
\vskip 0.4cm
$^{\rm a}$ARC Centre of Excellence for Particle Physics at the Terascale, \\ 
$^{\rm b}$School of Physics, The University of Sydney, 
NSW 2006, Australia 
\vspace{1cm}

\begin{abstract}
\noindent 
We describe a new phenomenon of zitterbewegung of a free Dirac particle in cosmological spacetimes.  Unlike the similar effect theorized by Schr\"odinger in 1930, the cosmological zitterbewegung is a real, physically attainable effect, which originates from the mixing of positive and negative frequency modes of a field operator in cosmological spacetimes. We briefly discuss the potential for observing this effect in laboratory experiments with trapped ions.

\end{abstract}

\end{center}

\vskip 3cm

\centerline {\bf Essay  received an Honorable Mention in the }
\centerline{\bf  Gravity Research Foundation essay contest 2013}

\vskip 1cm


\end{titlepage}


A consistent merger of relativity and quantum mechanics brings in full
flavour of interesting, qualitatively new physical phenomena. The great example of this is Dirac's relativistic 
theory of electron \cite{Dirac:1928hu}, which not only successfully explained the origin of the electron's spin and its magnetic moment, but also introduced the notion of antiparticles and laid  down a foundation for modern quantum-mechanical description of vacuum.     

It took quite sometime for these revolutionary insights to be properly understood within the framework of quantum field theory. In the early days, misinterpretations of negative frequency solutions of the Dirac equation often lead to physically controversial  conclusions. In 1930 E. Schr\"odinger published a paper \cite{zitter} where, by allowing a linear superposition of positive and negative frequency electron wavefunctions, he demonstrated that a free relativistic electron  undergoes a trembling motion. This phenomenon of zitterbewegung was in odds with Ehrenfest's theorem \cite{ehrenfest}, according to which the expectation value of the velocity operator of a free electron should behave as the classical velocity, i.e. it must be constant.   As it was understood subsequently, Schrodinger's derivation of the zitterbewegung effect was based on an incorrect treatment of negative frequency modes. With the correct approach both within the first-quantized Dirac equation in the Foldy-Wouthuysen representation \cite{Foldy:1949wa} and within the second-quantized theory  one can demonstrate that zitterbewegung is an unphysical effect, unobservable by any inertial observer.

Even more unusual phenomena emerge when one considers quantum fields in curved spacetimes, e.g. within the framework of semiclassical gravity, where  inertial reference frames are defined only locally. Since an inertial frame is not available globally, the separation of negative and positive frequency modes is typically impossible globally, and thus the standard flat spacetime notions of particle and anti-particle become ambiguous. This lead to a particle creation phenomenon in curved spacetime which is manifested in the Hawking radiation of black holes \cite{Hawking:1974rv}, the Unruh radiation in an uniformly accelerated reference frame \cite{Fulling:1972md}, the cosmological particle production \cite{Parker:1968mv},  which is believed to be responsible for primordial anisotropies in the cosmic microwave background radiation generated during inflation, and other related phenomena. 

In this essay we would like to describe a new phenomenon of cosmological zitterbewegung, which unlike the effect originally proposed by Schrodinger  in flat spacetime \cite{zitter}, is a physical effect precisely because of intrinsic mixing of positive and negative modes of a free Dirac field in cosmological spacetimes. Similar effect has been recently considered by one of us for the Dirac equation in an uniformly 
accelerated reference frame \cite{zitter1}.            

In what follows we consider a toy cosmological model in (1+1)-dimensions described by the metric:
\begin{equation}
ds^2=dt^2-a^2(t)dx^2~,
\label{1}
\end{equation}
where $a(t)$ is a scale factor that describes a given cosmological model. The essential physics of the cosmological zitterbewegung can be demonstrated using this simple (1+1)-dimensional model, without unnecessary technical complications.  We stress that the effect can be obtained in more realistic cosmological models, which we consider elsewhere. 

Dirac equation in the above cosmological spacetime (\ref{1}) can be written as:
\begin{equation}
i\partial_t \psi = \hat {\cal H} \psi~, 
\label{1a}
\end{equation}
where $\hat {\cal H}$ is the time-dependent Hamiltonian operator, 
\begin{equation}
\hat {\cal H}=-i\gamma^0\gamma^1\partial_x-\frac{i}{2}\partial_{t}\ln(a)+\gamma^0am, 
\label{2}
\end{equation} 
that describes evolution of the system in coordinate time $t$. Here $m$ is the mass of the particle and $\gamma^0, \gamma^1$ are 2-dimensional gamma matrices  Note, because of this time dependence the Hamiltonian $\hat {\cal H}$ (\ref{2}) cannot be diagonalized. Therefore, if we define instantaneous positive, $\psi_p^{(+)}$, and negative, $\psi_p^{(-)}$, frequency ($\omega_p=+\sqrt{p^2+m^2a^2}$) solutions to the Dirac equation (\ref{1a}), say at initial time $t_0$, 
\begin{equation}
  \psi_p^{(\pm)}=\frac{1}{\sqrt{2\omega_pV}}\left(
  \begin{tabular}{c}
 $ \sqrt{\omega_p\mp p} $\\ 
 $ \pm\sqrt{\omega_p\pm p} $\\ 
  \end{tabular} 
\right)  {\rm e}^{\mp i\int^{t_0} \omega_p d\tau +ipx}
 \label{3}    
\end{equation}
they will inevitably mix at a later time $t>t_0$.  Consequently, the quantized Dirac field is given in terms of these mixed modes:
\begin{eqnarray}
\hat \psi_t =\int dp\sqrt{\frac{V}{2\pi}}\left[\hat A(t,p)\left(\alpha(t)\psi^{(+)}_p+\beta(t)\psi^{(-)}_p\right)+\right. \nonumber \\\left.
\hat B^{\dagger}(t,-p)\left(\alpha^{*}(t)\psi^{(-)}_p+\beta^{*}(t)\psi^{(+)}_p\right) \right]~,  
\label{4}
\end{eqnarray}
where $A(t,p)$ ($A^{\dagger}(t,p)$) and $B(t,p)$ ($B^{\dagger}(t,p)$) are annihilation (creation) operators for particles and anti-particles, respectively, at an instant time $t$. They satisfy the standard anti-commutation relations. The complex mixing coefficients $\alpha(t)$ and $\beta(t)$ satisfy the normalization equation $\vert \alpha \vert^2+\vert \beta\vert^2=1$. They are known as Bogoliubov coefficients.

Within the above second-quantized formalism it is straightforward to compute a wave-function corresponding to a one-particle state $\vert p,t\rangle$ defined at a given instant $t$ and carrying momentum $p$:
\begin{equation}
\phi_p=\left(\frac{2\pi}{V}\right)^{1/2} \langle 0,t\vert \hat \psi_t\vert p,t\rangle=\alpha(t)\psi^{(+)}_p+\beta(t)\psi^{(-)}_p~,
\label{5}
\end{equation}       
where $\vert 0,t\rangle$ is an instantaneous vacuum state, $\hat A(t,p)\vert 0,t\rangle=\hat B(t,p)\vert 0,t\rangle=0,~\forall p$.  As it was anticipated, one-particle wavefunction at a time $t>t_0$ (\ref{5}) is a mixture of positive and negative frequency modes, even though one started with only positive frequency wavefunction at instant $t_0$, $\alpha(t_0)=1, \beta(t_0)=0$. 

Next we define the coordinate velocity operator, 
\begin{equation}
\hat v=-i\left[x, \hat {\cal H}\right]=\gamma^0\gamma^1~,
\label{6}
\end{equation} 
which has the same form as the one in flat spacetime. This operator is defined in the Schr\"odinger picture, as it does not depend on time.  We then compute its expectation value in one-particle state (\ref{5}):
\begin{eqnarray}
v=\int a dx\left[\phi^{\dagger}_p\gamma^0\gamma^1 \phi_p\right]= \nonumber \\
\left(\vert\alpha\vert^2-\vert\beta\vert^2\right)\frac{p}{\omega_p}-\frac{ma}{\omega_p}\left[
\alpha^*\beta {\rm e}^{2i\int^{t_0}\omega_pd\tau}+c.c. \right]
\label{7}
\end{eqnarray}

The second term in the last line of the above equation is the cosmological zitterbewegung term, which is non-vanishing even in the reference frame where particle is at rest, $p=0$. It vanishes in flat space time ($a(t)=1$), since  $\alpha(t)=1$ and $\beta(t)=0$ for $\forall t$.  The velocity in this case becomes $v=\frac{p}{\omega_p}$, as it should be for a free relativistic particle with momentum $p$ and energy $\omega_p=\sqrt{p^2+m^2}$ measured in a given inertial reference frame. Thus, unlike Schr\"odinger's original proposal \cite{zitter}, the cosmological  zitterbewegung described in this essay is a real physical effect. It may have interesting cosmological implications. A similar effect can be found in an uniformly accelerated reference frame and black hole spacetimes \cite{zitter1}.

Apart from the purely theoretical interest related to a new unusual manifestation of quantum fields in curved spacetimes, the cosmological zitterbewegung effect can be a subject of potential experimental studies. Namely, there are suggestions to study various quantum effects in cosmological spacetimes by means of suitable laboratory analogues (see, e.g., \cite{Barcelo:2005fc} for a recent review). In this regard, it would be interesting to investigate whether it is feasible to observe cosmological zitterbewegung in such laboratory experiments. Without going into details, we intuitively expect that the cosmological expansion (contraction) can be mimicked by an expanding (contracting)  medium. The obvious challenges for such experiments are: (i)~ The temperature must be kept low enough, so that thermal fluctuations do not overshadow the quantum zitterbewegung effect, and (ii)~The generated time-dependence must be such that, that the effective dynamics is described by eq. (\ref{1a}), and it also must be sufficiently strong to produce observable zitterbewegung.  Because of low temperatures, trapped ions may constitute a promising experimental set-up for the observation of the cosmological zitterbewegung effect, providing rapid and controlled expansion/contraction of the ion trap can be achieved \cite{Schutzhold:2007mx}. 

To conclude, we have described in this essay a new physical effect of zitterbewegung of a free Dirac particle in cosmological spacetimes and suggest its possible experimental verification in laboratory experiments with trapped ions.   



\end{document}